\documentclass[sigconf]{acmart}
\renewcommand\footnotetextcopyrightpermission[1]{}
\usepackage{booktabs} 
\usepackage{flexisym}

\usepackage{balance} 

\acmConference[RecSys 2017 Poster Proceedings]{RecSys 2017 Posters}{August 27-31}{Como, Italy} 

\begin{document}
\title{Towards Effective Exploration/Exploitation in \\ Sequential Music Recommendation}

\author{Himan Abdollahpouri}
\affiliation{%
  \institution{DePaul University}
  \country{USA}
}
\email{habdolla@depaul.edu}
\author{Steve Essinger}
\affiliation{%
  \institution{Pandora Media, Inc.}
  \country{USA} 
}
\email{sessinger@pandora.com}

\begin{abstract}
Music streaming companies collectively serve billions of songs per day. Radio-based music services may intersperse audio advertisements among the songs as a means to generate revenue, much like traditional FM radio. Regardless of the monetization  approach, the recommender system should decide when to play content that the listener is known to enjoy (exploit) and content that is novel to the listener (explore). Recommender systems that rely on this explore/exploit type framework have been deployed in a wide variety of applications such as movies, books, music, shopping and more. In this work, we investigate the impact of different ad/song sequences on listener behavior. In particular, we focus on the impact of exploring new song content for the listener given the previous sequence of ads and songs in the listener's session. Our results show that the prior sequence matters when considering song exploration and that this prior sequence has an impact on the listener's tendency to interrupt their current session.
\end{abstract}

%
%

\keywords{}

\maketitle

\section{Introduction}
Recommender systems (RS) have been deployed in numerous domains including music, movies, e-commerce and books. In music recommendation, one of the overarching goals of the RS is to find the best song to play for each listener, personalized to their specific taste(s) in music. In general, companies offering music recommendation services provide two different types of subscriptions: (1) Ad-supported membership where the music is free, but the listener is subject to advertisements and (2) premium membership where listener pays a monthly membership fee in exchange for ad-free listening. This paper focuses on the former, ad-supported listening. Unsurprisingly, listeners prefer hearing songs over ads. However, the business depends on the revenue that it makes from the ads and cannot operate without serving them. Therefore, playing ads is crucial to keep the business alive and should be considered as a content served to the listener along with music.
{\let\thefootnote\relax\footnote{RecSys 2017 Poster Proceedings, August 27-31, Como, Italy.}}

One of the fundamental concepts in RS is the idea of exploration and exploitation \cite{ vanchinathan2014explore}. This paradigm results in a balance between recommending content the system has high certainty the user would like (exploitation) and the content for which there is less certainty (exploration). Without exploration, users would become stuck in a filter bubble and continue to see a narrow set of products. This is a missed opportunity to experience other products that could be of interest to them \cite{resnick2013bursting, celma2016exploit}. Another reason for exploration is when the number of items matching a user's interest is limited and the system should not recommend the same item again to the user. For example, in online dating \cite{reciprocal}, it is possible that the system has already recommended all the available people who match the user's interest and, so, exploring a wider range of people is needed in order to be able to generate new recommendations. Therefore, providing exploratory content to a user is a key component for discovery. We conducted an experiment on a music recommendation application and our results show that the previous sequence of events in a listener's session is important in deciding whether the RS should provide subsequent exploratory types of content.

\section{Song/Ad Sequence Analysis}

\begin{figure}[b]
    \centering
    \includegraphics[height=2.2in, width=3.2in]{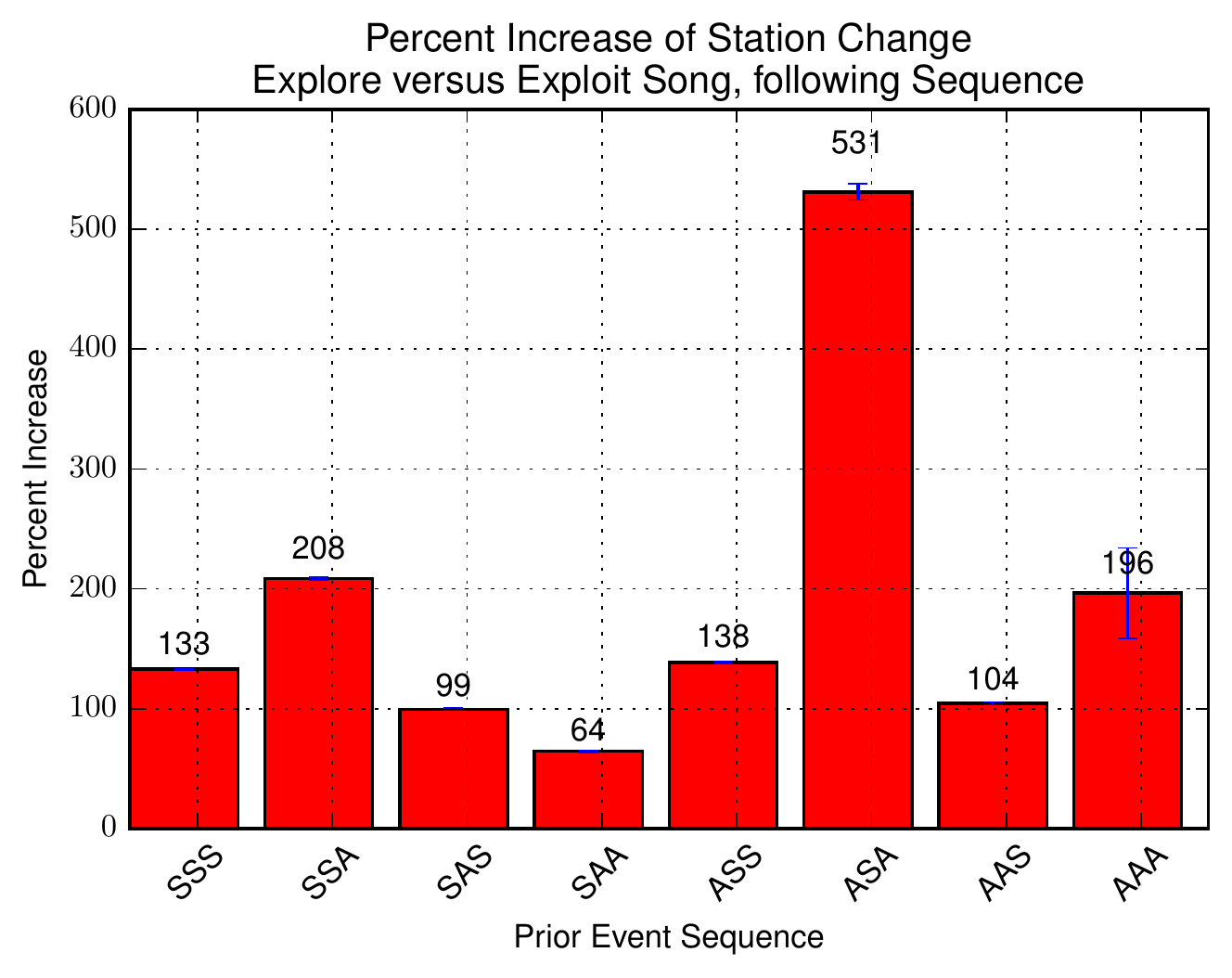}
    \caption{Percent increase of the probability of station change for an explore song vs. an exploit song, following different sequences of exploit songs and ads.}
    \label{fig:lt-effect}
\end{figure}
We have compiled data from a large-scale music recommendation service for our analysis. To find the effect of different sequences of songs and ads on the probability of a user switching the station after listening to an exploratory song, we looked at one million sessions on mobile devices where the ad placement had been made completely at random. Note that the randomness of ad placement is important in order to make sure our analysis is not biased toward any particular ad placement algorithm. We compare the impact of explore songs versus exploit songs in the context of the previous three events. For example, given the prior three events Ad, Song, Song, where each song is an exploit, what is the probability of the listener changing the station if the next song spun for them is an explore song versus the probability of station change given an exploit song? Station change is used as a proxy for discontent with the current stream of music. 

We calculated the probabilities of users changing the station when they are exposed to different sequences of ads and songs as follows: there are a total of 8 possible event combinations for a set of three items as shown in figure ~\ref{fig:lt-effect}. We denote explore song by \textit{S\textprime} and exploit song by \textit{S}. \textit{Station change} is represented by, \textit{C}.
\begin{equation}
   \text{Percentage Difference} \quad = \quad \frac{P(C\; |\; S\textprime) - P(C\; |\; S)}{P(C\; |\; S)}*100
\end{equation}
$P(C\; |\; S\textprime)$ is the probability of a user changing the station given the last played content is an explore song. $P(C\; |\; S)$ is the probability of a user changing the station when the last played content is an exploit song.
The lower and upper confidence bounds for the computed percentage increases, shown in figure ~\ref{fig:lt-effect} as vertical blue lines on top of the bars, are computed as follows,

\begin{equation}
 \left[\frac{P(C\; |\; S\textprime) - P(C\; |\; S)}{P(C \;|\; S)} \pm 1.96*SE \right] * 100
\end{equation}

where SE (i.e. the standard error) is calculated using equation 3,

\begin{equation}
\sqrt{\frac{P(C\; |\;S\textprime )*(1-P(C\; |\; S\textprime))}{N_{S\textprime}} + \frac{P(C \;|\; S)*(1-P(C \;|\; S))}{N_{S}}}
\end{equation}

where $N_{S\textprime}$ is the total number of times an explore song has been played. The total number of times an exploit song has been played is denoted by $N_{S}$.
Figure ~\ref{fig:lt-effect} shows the percent increase of station changes after playing an explore versus an exploit song when a user has observed the respective prior sequence of exploit songs and ads. Due to the company's data privacy policy, we have not included the individual probabilities of switching the station for explore and exploit songs, but have provided the probability difference of change. 

An exploit song is denoted by \textit{S} and an ad is shown by \textit{A}. As you can see, depending upon the previous sequence of songs and ads, the probability of a user switching the station when we show them an explore song is higher than the probability when we show an exploit song. This is true for all 8 different combinations of songs and ads. Moreover, some sequences are riskier than the others for placing an explore song. For example, the \textit{ASA} sequence (which means playing an ad, then a song and then another ad) has the highest probability increase (+531.13\%) of a user switching the station when given an explore song after that sequence. Clearly, this is not the best opportunity to explore new content. On the other hand, the \textit{SAA} sequence has the lowest probability increase (+64.42\%), but is still positive. While playing an explore song is still riskier than an exploit song in all cases, it is better to explore after particular sequences over others. Certainly, different sequences of songs and ads have different effects on station switching behavior and a recommender system should try to take these sequences into account when doing exploration and exploitation, as in our sequential music recommendation system. Overarching, instead of a blind explore-exploit platform, we advise taking an intelligent approach that accounts for a listener's state of listening (whether they are happy with the past couple of songs/ads or not) into account when deciding to exploit or when to explore. 

\section{Related Work}
The idea of explore-exploit has been studied in recommender systems by some researchers \cite{vanchinathan2014explore}. In particular, for single item recommendation, approaches like Multi-Armed Bandits have been used to make a balance between exploration and exploitation \cite{wang2014exploration}. Moreover, authors have previously proposed an approach for an effective balance between recommending popular and long-tail items \cite{abdollahpouri2017controlling}. A more similar idea to our work is done in \cite{dali2015please} where authors investigated a proper timing for delivering the recommendation. However, in our work, we are not looking for a perfect timing for the recommendation in general as the user always should receive a content (song or ad) as recommendation. Our work is also novel as we look at the previous sequences of the recommendations as an indication for whether it is a good time for exploration or not.

\section{Conclusion and Future Work}
In this work, we investigated the impact of different ad/song sequences on listener behavior. In particular, we focused on the impact of exploring new song content for the listener given the previous set of ads and songs in the listener's session. Our experimental results show that the previous sequence of ads/songs matters in deciding what the right time is for exploration versus exploitation. For our future work, we will launch an A/B experiment controlling for the placement of explore songs and see how different users behave when they observe different sequences of songs and ads. We will also investigate more sophisticated offline models, such as HMMs and RNNs in a reinforcement learning setting that could learn superior personalized playlist sequencing. This work is a starting point for a larger project in which we aim to optimize the stream of recommendations of mixed types of content (i.e. contents from different stakeholders) \cite{vamsBurkeHiman, umapHimanMS, vamsSteveHiman}. 
\begin{acks}
We would like to thank Pandora Media, Inc. for access to their vastly rich dataset.
\end{acks}

\balance

\end{document}